# Fully Distributed Secondary Voltage Control in Inverter-Based Microgrids

Ali Dehghan Banadaki[1], *Student Member, IEEE*, Ali Feliachi[1], *Senior Member, IEEE,*
Vinod K. Kulathumani[2], *Senior Member, IEEE*
Lane Department of Computer Science and Electrical Engineering,
West Virginia University, Morgantown, WV 26505, USA

*Abstract*— Centralized secondary voltage control in a power system has been replaced by the distributed controller in the recent literature due to its high dependency on extensive communication messages. Although in the new method each distributed generator only communicate with its neighbors to control the voltage, yet the messages are circulating among the whole system. In this paper, we have utilized distributed controller locally so that it will work as a fully distributed control system. This controller has been justified by being studied within a case study including 6 distributed generators.

## I. Introduction:

Microgrid (MG) penetration in a power system is expanding in recent years due to its higher reliability and more energy efficiency. Utilizing renewable energies such as solar power, wind power, and fuel cells will save a huge amount of money as well as reducing the carbon dioxide emission into the air.

Due to the alternative nature of renewable energies, power electronic devices are commonly used as an interface between the renewable energies and the grid [1]. Controlling these distributed generators (DGs) unlike the synchronous machines in conventional power systems would be different due to the lack of inertia.

MGs can operate in both islanded mode and grid-connected mode [2]. In the latter case, voltage and frequency will be regulated by the main grid while in the former case, DGs should take control of voltage and frequency. In synchronous machines, the inertia of the rotor is much greater than the DGs. Therefore, once the load is changed in the system, mechanical power of the rotor will be transformed to the electrical power to be injected into the power system. Due to the null or much lower inertia in the DGs, this behavior has been mimicked by using droop equations. Voltage Control Voltage Source Inverter (VCVSI) which is a common DG source, has been implemented with droop equations in this paper [3].

A typical voltage control in MG can be categorized into two levels, namely, primary and secondary. Although primary control level can stabilize the system at the first point the voltage regulation needs to be adjusted by the secondary controller. In the secondary controller, researchers have used a centralized algorithm to achieve the global minimum [4]. However, due to the high dependency of this algorithm to the communication network, the probability of single point of failure is very high. Hence, the distributed algorithm has been proposed in [5] to alleviate single point of failure. In these algorithms, DGs communicate together to bring back their voltages to the reference or the average value. In other words, neighboring DGs will communicate together to reach an agreement that can be either a reference value or an average value. This method is called distributed voltage consensus. While in this method, each DG will communicate only with its neighboring DGs, eventually the global messaging happens among them.

In this paper, we insist on the fact that the voltage control is a local phenomenon rather than a global one. Therefore, a real distributed algorithm should only work on the regions that can have maximum effect on the area around the disturbance. Therefore in this paper, DGs that are in the close vicinity of the disturbance and not just neighboring in terms of the communication network will be used to control the system.

Rest of this paper is organized as follows: In section II, an introduction to the structure of DG that has been modeled in the system is given. In section III, distributed control algorithm will be investigated. In section IV, a case study to show the importance of this algorithm is shown and finally in section V conclusion is given.

## II. Voltage Control Voltage Source Inverters:

Block Diagram of a VCVSI is shown in Figure (1). It consists of AC/DC/AC converter, LC filter, and an output connector. It also consists of three controlling parts including a current controller, a voltage controller, and a power controller. In the power controller, two equations namely frequency droop and voltage droop are being used. In the voltage controller, the actual voltage and the output voltage are compared to set a reference point for the current controller.

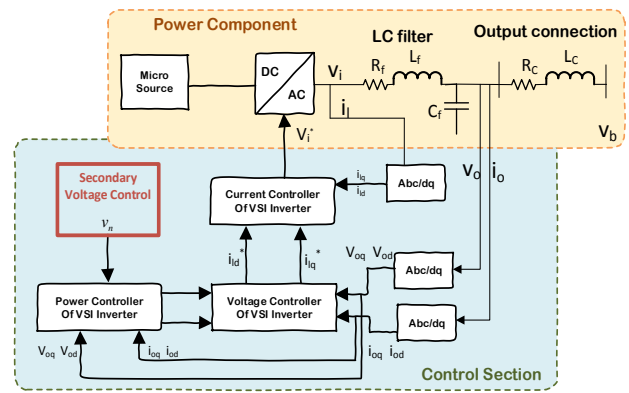

Figure 1: Block Diagram of VCVSI

[1]Advanced Power & Electricity Research Group (APERC)
[2]Wireless Sensor Actuator Networks (WSAN) Research Group

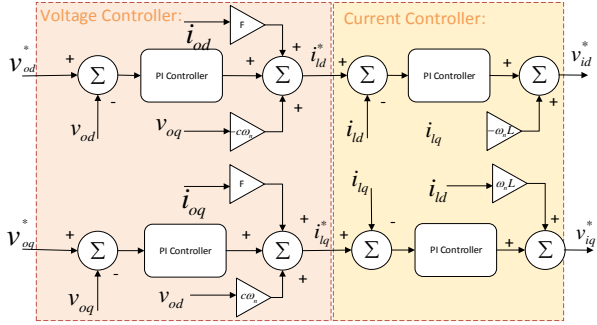

Figure 2. Voltage and Current Controller

Finally in the current controller, the actual current and the reference current from the voltage controller will be compared to finally set the magnitude of the output voltage that the converter should follow. On the other hand, the frequency of the converter will be adjusted by the droop frequency in the power controller. Explanation about each part of controller of a VCVSI is given in the following:

A. *Power controller:*

Voltage and frequency droop equations are utilized as a negative feedback controller in power controller. Their task is to reduce the voltage/frequency magnitude in terms of increasement in reactive/active power. The power controller block has three parts that are explained below:

*1)* Power calculation block calculates the output power in the system via (1,2):

$$p = v_{od}i_{od} + v_{oq}i_{oq} \quad (1)$$
$$q = v_{od}i_{oq} - v_{oq}i_{od} \quad (2)$$

*2)* Low pass filter: a low pass filter shown in equations (3, 4), is used in the power controller to filter the high switching frequencies (harmonics) coming from the switching of inverter-based MGs:

$$P = (\omega_{co}/(s+\omega_{co}))p \quad (3)$$
$$Q = (\omega_{co}/(s+\omega_{co}))q \quad (4)$$

*3)* Controller: Primary controller implements two droop equations for sharing active and reactive power based on the rating of the sources in MGs:

$$\omega = \omega_n - mP \quad (5)$$
$$v^*_{od} = V_n - nQ \quad (6)$$
$$v^*_{oq} = 0 \quad (7)$$

Where m and n are the droop gains given in (8, 9):

$$m = \frac{\Delta \omega_{max}}{P_{max}} \quad n = \frac{\Delta V_{od\,max}}{Q_{max}} \quad (8, 9)$$

B. *Voltage controller:* A PI controller with considering the feedbacks and feedforwards consists the voltage controller shown in Fig. (2) and formulated in (10-12).

$$\dot{\phi}_{dq} = v^*_{odq} - v_{odq} \quad (10)$$
$$i^*_{ld} = Fi_{od} - \omega_n C_f v_{oq} + K_{pv}(v^*_{od} - v_{od}) + K_{iv}\phi_d \quad (11)$$
$$i^*_{lq} = Fi_{oq} + \omega_n C_f v_{od} + K_{pv}(v^*_{oq} - v_{oq}) + K_{iv}\phi_q \quad (12)$$

C. *Current Controller:* A PI Current controller is used as the for adjusting the set points of input voltage references:

$$\dot{\gamma}_{dq} = i^*_{ldq} - i_{ldq} \quad (13)$$
$$v^*_{id} = -\omega_n L_f i_{lq} + K_{pc}(i^*_{ld} - i_{ld}) + K_{ic}\gamma_d \quad (14)$$
$$v^*_{iq} = -\omega_n L_f i_{ld} + K_{pc}(i^*_{lq} - i_{lq}) + K_{ic}\gamma_q \quad (15)$$

D. *Output RLC filter and Coupling inductor:* D-Q equations of output filter and coupling inductor are given in (16-21):

$$\frac{di_{ld}}{dt} = \frac{-r_f}{L_f}i_{ld} - \omega i_{lq} + \frac{1}{L_f}v_{id} - \frac{1}{L_f}v_{od} \quad (16)$$
$$\frac{di_{lq}}{dt} = \frac{-r_f}{L_f}i_{lq} - \omega i_{ld} + \frac{1}{L_f}v_{iq} - \frac{1}{L_f}v_{oq} \quad (17)$$
$$\frac{dv_{od}}{dt} = \omega v_{oq} + \frac{1}{C_f}i_{ld} - \frac{1}{C_f}i_{od} \quad (18)$$
$$\frac{dv_{oq}}{dt} = -\omega v_{od} + \frac{1}{C_f}i_{lq} - \frac{1}{C_f}i_{oq} \quad (19)$$
$$\frac{di_{od}}{dt} = \frac{-r_c}{L_c}i_{od} + \omega i_{oq} + \frac{1}{L_c}v_{od} - \frac{1}{L_c}v_{bd} \quad (20)$$
$$\frac{di_{oq}}{dt} = \frac{-r_c}{L_c}i_{oq} - \omega i_{od} + \frac{1}{L_c}v_{oq} - \frac{1}{L_c}v_{bq} \quad (21)$$

We have built our VCSVCI block in Matlab Simulink by implementing equations (1-21) from [6]. This DG is used for analyzing the zonal analysis in distributed voltage control in the next session.

### III. DISTRIBUTED SECONDARY VOLTAGE CONTROL:

Although a primary control level may stabilize the system without any communications, it might not be able to fully alleviate the voltage deviation in power systems. Therefore, the secondary controller readjusts the voltage set-points to control the MG (i.e. $v_n$ in Figure (1)). In contrast to the centralized algorithm, a distributed algorithm has been proposed in the literature. Although each DG only communicates with its immediate neighbors, the final control is done globally. Moreover, it might not even reach the desired reference value. Therefore, in this paper, we have considered the controlling part in a zonal effect among the neighboring DGs. In this section, the zonal effect of DGs for implementing real distributed control algorithm is discussed. It is a well-known fact that controlling a bus voltage will not be feasible from a far distant area. For example, load number 4 in a power system shown in Figure (3), can be controlled by DGs 4, 5 and 6 rather than DGs 1, 2, and 3.

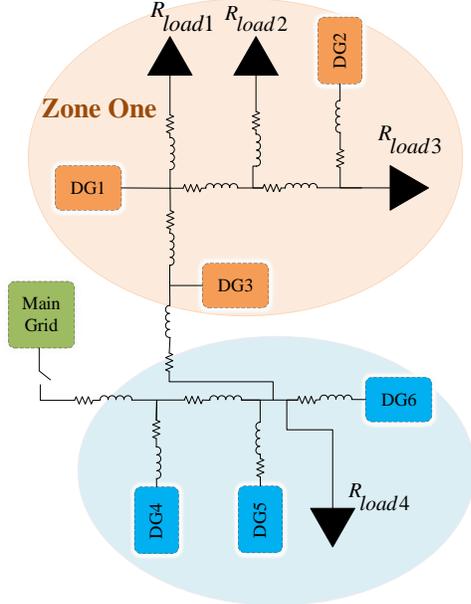

Figure 3. Zonal Analysis for Voltage Control

*A. consensus algorithm overview:*

Consensus algorithm is as a distributed algorithm that can reach an agreement by sharing each node's value with their neighbors. Although this algorithm tries to make an agreement among all of the neighbors, this should not be our intention in a voltage control. A distributed cooperative tracking algorithm is proposed in [5]. In this algorithm, one of the DGs is selected as a leader node. This DG knows the reference value which will be communicated to the other DGs via the communication graphs. Follower DGs will take that value to readjust their voltage controller set-points (i.e. $v_n$) via equation 22. In this algorithm, all the DGs should be at least connected to the leader DG via a connected graph. While this algorithm works better in systems that DGs are close to each other, we have extended this algorithm by considering zonal effects. Our communication network is shown in Figure 5 part 2 in contrast to part 1.

$$\Delta v_{ni} = c_{vi} \int (\sum_j a_{ij}(v_{odi} - v_{odj}) + g_i(v_{odi} - v_{ref}))dt \quad (22)$$

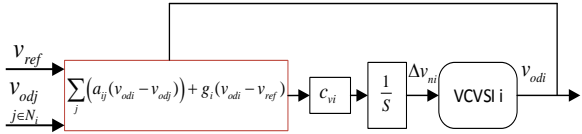

Figure 4: Distributed secondary voltage control

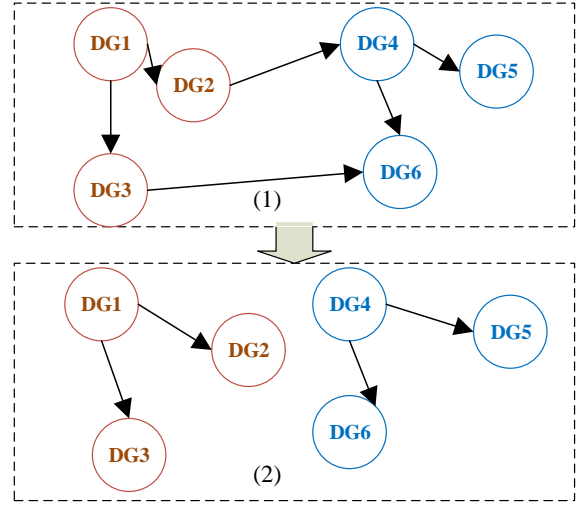

Figure 5. Two different DGs' Wireless Connection
(1) Global consensus connection;
(2) Zonal consensus connection

TABLE I: UNDERSTUDIED SYSTEM PARAMETERS

| | | | | | | | | |
|---|---|---|---|---|---|---|---|---|
| $r_f$ | 0.1 Ω | $L_c$ | 0.35 mH | $K_{pc}$ | 10.5 | $m_p$ | 9.4e-5 |
| $L_f$ | 1.35 mH | $\omega_c$ | 31.41 | $K_{ic}$ | 16e3 | $n_q$ | 1.3e-3 |
| $C_f$ | 50 μF | $r_c$ | 0.03 Ω | $K_{pv}$ | 0.05 | $K_{iv}$ | 390 |
| $c_v$ | [30 30 30] | $g_1$ | 1 | $a_{12}$ | 1 | $a_{13}$ | 1 |

*B. Simulation results:*

A power system with 6 DGs and 4 loads is studied here (Figure 3). DGs have been modeled in Matlab software by implementing the equations of section II. Parameter details of the system are shown in Table I. Here we consider two scenarios to show the effectiveness of our method:

*a) Scenario one:* the zonal division between two zones has been done as it is shown in Figure 3 and the communication between DGs are shown in Figure 5 part 2. The fully distributed secondary controller is activated at t=0.6 seconds to show how it can improve the voltage deviation after the disturbance. The results in Figures 6 and 7 show that the voltages of each group successfully converge to its reference value which is chosen to be 381.

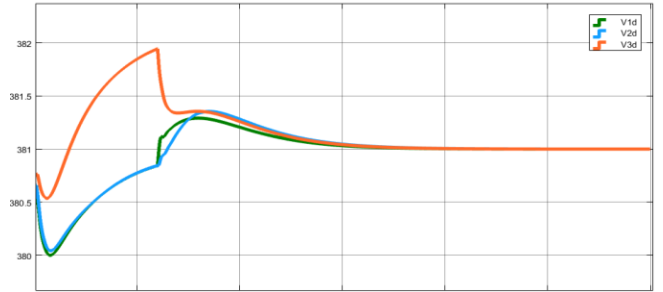

Figure 6. Voltages of DGs in zone one

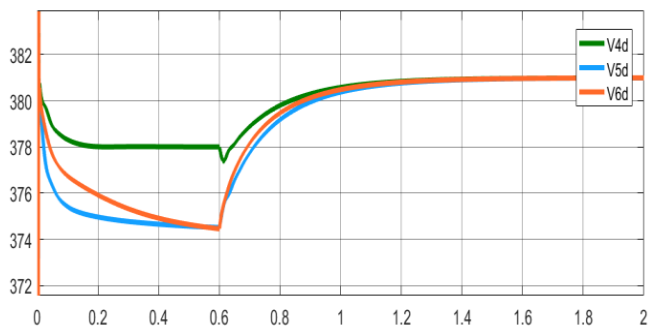

Figure 7. Voltages of DGs in zone two

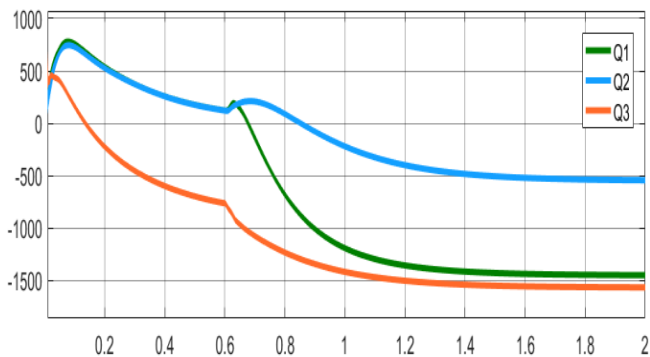

Figure 8. Reactive powers of DGs in zone one

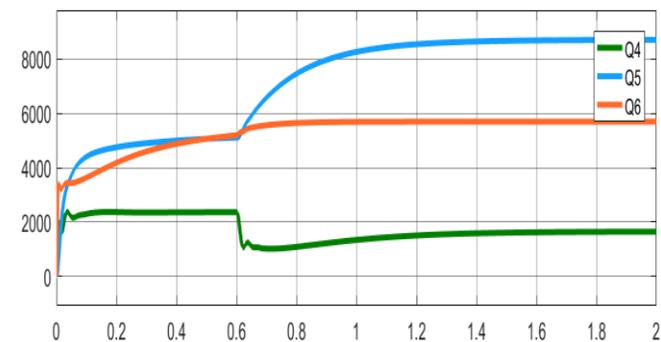

Figure 9. Reactive powers of DGs in zone two

*b) Scenario two:* In this scenario, voltage zoning is not applied to the system, therefore, communication of DGs is done as it is shown in Figure 5 part 1. DG1 is the leader here, unlike the previous part that we had two leaders for each zone (DG1 for zone 1 and DG4 for zone 2). Figure 10 shows that although the voltages still reach to their reference value, it took almost twice longer as it was the case for the previous one.

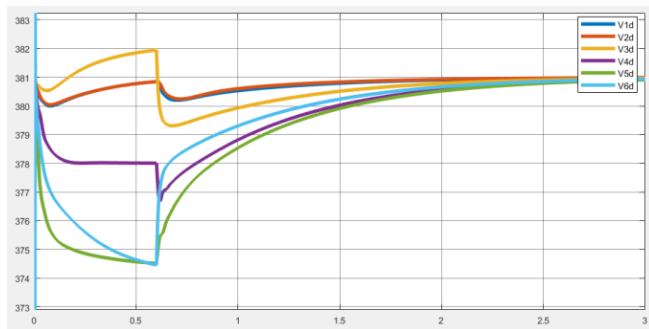

Figure 10. Voltages of DG1-6 in the second scenario

IV. CONCLUSION

In this paper, the fully distributed secondary voltage control has been proposed for microgrids which has been accomplished by considering the zonal effect in applying the consensus method. Our case study considers 6 DGs with 4 loads. DGs were deterministically placed in two clusters to prove that the voltages of DGs by considering zonal effect will reach to the reference value in a shorter time with fewer communications.